\documentclass[10pt,twocolumn]{article}
\usepackage{latexsym}
\usepackage{amsmath}
\usepackage[dvips]{graphicx}
\usepackage{fancyhdr}

\begin{document}

\newcommand{\ad}{a^{\dagger}}
\newcommand{\add}{a^{\dagger 2}}
\newcommand{\addd}{a^{\dagger 3}}
\newcommand{\adddd}{a^{\dagger 4}}
\newcommand{\fd}{f^{\dagger}}

\title{Detailed study of a transition point \\ in the Veneziano-Wosiek model of Planar Quantum Mechanics}
\author{Piotr Korcyl \\ \small{\emph{M. Smoluchowski Institute of Physics, Jagiellonian
University}} \\ \small{\emph{Reymonta 4, 30-059 Krak\'{o}w,
Poland}}}
\date{}

\twocolumn[
\begin{@twocolumnfalse}
      \maketitle
      \begin{abstract}
        Following a model recently investigated by Veneziano
        and Wosiek we briefly introduce Planar Quantum Mechanics (PQM).
        Then, we present high precision numerical
        results in the sectors with two and three fermions. We
        confirm, that the transition
        point in the 't Hooft's coupling constant $\lambda$ in
        these sectors occurs at
        $\lambda_c = 1$, as was expected.\\\\
      \end{abstract}
    \end{@twocolumnfalse}

    \vspace{10cm}

    \begin{flushleft}
    TPJU- \\ 2006
    \end{flushleft}
]

\pagebreak

    The exact numerical spectra of a branch of reduced supersymmetric theories can be
    calculated  in a cut Fock basis by a method proposed recently by \mbox{Wosiek \cite{wosiek}}. In a
    series of papers \cite{cw2, cw3} he studied
    some models of Supersymmetric \mbox{Yang-Mills} Quantum
    Mechanics (SYMQM). These systems
    result from a dimensional reduction of the full dimensional
    \mbox{(D = d + 1)} supersymmetric Yang-Mills quantum field theories to a single point in
    space (0 + 1). The method provided a good understanding of
    the \mbox{D = 2} and D = 4, $N$ = 2 spectra \cite{cw2, cw3}. The goal of
    such analysis is to reach, on one
    hand the system with \mbox{D = 4}, \mbox{$N$ = 3} which could shade some light
    on the real QCD, and on the other, the D = 10, SU($N \rightarrow \infty$)
    model, which is
    conjectured to be in relation with the M-theory \cite{susskind}.
    The latter case needs to incorporate into the scheme the \mbox{large $N$ limit
    \cite{thooft}}, which is expected to provide a remarkable simplification.
    It should reduce considerably the number of basis vectors to be taken
    into account and allow to disregard all non-planar contributions.
    The above was investigated by Veneziano and Wosiek in
    \cite{vw1, vw2, vw3} on a simple supersymmetric model belonging to
    the class of Planar Quantum Mechanics (PQM).

    In this paper we present high precision numerical results on the
    model from Ref\cite{vw1} in the sectors with two and three
    fermions. We investigate the transition point in the 't Hooft's
    coupling $\lambda$, and show that it indeed takes place at
    $\lambda_c = 1$. Therefore we give a numerical confirmation of
    the analytical results obtained by Beccaria in \cite{beccaria}.

    The paper is constructed as follows. We start with an
    introductory part to the PQM, then we present the studied model,
    and finally we discuss our results.

    \subsubsection*{Supersymmetric Yang-Mills \\ Quantum Mechanics}

    We will introduce now the systems called SYMQM in the Hamiltonian
    formulation. Let's consider a quantum mechanical system with
    $N^2$ bosonic and $N^2$ fermionic degrees of freedom. As was
    already
    mentioned, it can be regarded as the remainder after a dimensional
    reduction of supersymmetric field theory with U($N$) gauge
    symmetry to one point in space. During such a procedure, a local
    gauge symmetry becomes a global one. Thus, our system should be
    invariant under a global U($N$) rotation. Let $T^a_{ij}$ be the
    generators of the U($N$) group in the fundamental
    representation, thus, they are $N \times N$ matrices. We
    introduce bosonic and fermionic matrix-valued annihilation and creation
    operators
    \begin{equation}
    a_{ij} = a^b T^b_{ij}, \quad a^{\dagger}_{ij} = a^{\dagger b}
    T^b_{ij},
    \end{equation}
    \begin{equation}
    f_{ij} = f^b T^b_{ij}, \quad f^{\dagger}_{ij} = f^{\dagger b}
    T^b_{ij},
    \end{equation}
    where the sum over $b = 1, \dots, N^2$ is assumed, and $i,j = 1,
    \dots, N$. The invariance of the system is assured by taking the
    Hamiltonian as a trace of a polynomial of the above operators.
    The creation and annihilation operators satisfy the following
    commutation and anticommutation relations
    \begin{equation}
    [a_{ij}, a^{\dagger}_{kl}] = \delta_{il}\delta_{jk},
    \end{equation}
    \begin{equation}
    \{f_{ij}, f^{\dagger}_{kl}\} = \delta_{il}\delta_{jk}.
    \end{equation}
    The Fock basis is composed of eigenstates of the occupation
    number operators, $B = Tr(a^{\dagger}a)$ and $F =
    Tr(f^{\dagger}f)$, which are explicitly U($N$)-invariant. The
    construction of the basis starts from the Fock \mbox{vacuum denoted by $|0>$}. We act on
    the latter with invariant 'bricks' i.e. creation operators contracted with U($N$) invariant tensors.
    For the U($2$) group we have
    two such tensors, the $\delta_{ij}$ and $\epsilon_{ijk}$. The
    basis states are obtained by an action of any combination of
    powers of theses bricks.

    \subsubsection*{The cut-off method}

    As it is impossible to handle infinite matrices on a PC, one
    needs to cut them somehow. The most intuitive way to do this is
    to introduce some integer, $B_{max}$, and to keep only those
    basis states for which the total bosonic occupation number does
    not exceed $B_{max}$. The method of obtaining the spectrum
    simply consists of calculation of the hamiltonian matrix
    elements in such a cut Fock basis, and then of its numerical
    diagonalization. This should be done for several different
    cut-offs and a limit of infinite cut-off should be extrapolated
    in order to obtain a $B_{max}$ independent, thus physical,
    results. The difficulty of such a program is hidden in the
    number of basis states growing exponentially with increasing
    $B_{max}$ and $N$. Up to now, calculations have been made up to
    U($4$) \cite{pr_cam}.

    \subsubsection*{Large $N$ limit and Planar Quantum Mechanics}

    The difficulties described in the preceding paragraph largely disappear
    in the 't Hooft limit, $N \rightarrow \infty$, $g^2 N = \textrm{const}$,
    where $g$ is a coupling constant present in the system. The zeroth order
    approximation in the $\frac{1}{N}$ expansion consists in
    retaining only those contributions which correspond to planar
    graphs. It appears that this can be done already on the level of
    the Fock basis \cite{vw1}. The main contribution will be given
    by basis states obtained by action of a single trace brick.
    Therefore, in the purely bosonic sector, $F=0$, one needs to
    consider only basis states of the form
    \begin{equation}
    |0,n> = \frac{1}{\mathcal{N}_{0,n}}Tr[(a^{\dagger})^n]|0>,
    \end{equation}
    which are labeled by one integer, $n$. We can calculate an
    explicit expression for the normalization constant $\mathcal{N}_{0,n}$ \cite{vw1}.
    The sector $F=1$ contains one fermion and the basis states are given by
    \begin{equation}
    |1,n> =
    \frac{1}{\mathcal{N}_{1,n}}Tr[(a^{\dagger})^{n}f^{\dagger}]|0>.
    \end{equation}
    With increasing fermionic occupation number $F$, things get
    complicated, because one has to use several integers to label
    basis states. For example, if $F=2$ we need two integers, $n_1$ and $n_2$.
    The basis state is therefore obtained by the action of a trace
    \cite{vw2}
    \begin{equation}
    |2, n_1, n_2> = \frac{1}{\mathcal{N}_{2,n_1, n_2}}
    Tr[(a^{\dagger})^{n_1}f^{\dagger}(a^{\dagger})^{n_2}f^{\dagger}]|0>.
    \end{equation}
    Due to the cyclicity of the trace, we only need to deal with
    states with $n_1 \le n_2$. Moreover, if $n_1 = n_2 = m$, the
    anticommutation of fermionic creation operators and the
    cyclicity of the trace imply that \mbox{$|2, m, m> = 0$}. So, the
    basis is composed of states for which \mbox{$n_1 < n_2$}. Thus, for
    a given cut-off $B_{max}$, we will have
    $\frac{1}{2}B_{max}(B_{max}-1)$ states.

    In the case of three fermions, we need three integers to
    label a basis state \cite{vw2}
    \begin{eqnarray}
    &&|3,n_1, n_2, n_3> = \\
    && =\frac{1}{\mathcal{N}_{2,n_1, n_2, n_3}} Tr[\fd (\ad)^{n_1} \fd
(\ad)^{n_2} \fd
    (\ad)^{n_3}]|0>. \nonumber
    \end{eqnarray}
    Again, we can arrange them so that $n_1 < n_2, n_3$.

    The large $N$ limit of SYMQM systems in the zeroth order
    approximation is called Planar Quantum Mechanics.

    \subsection*{Veneziano-Wosiek model}

    The model considered in Ref\cite{vw1,vw2,vw3} is given by the
    supersymmetry generators
    \begin{equation}
    Q = Tr[fa^{\dagger}(1+ga^{\dagger})], \quad Q^{\dagger} =
    Tr[f^{\dagger}(1+ga)a],
    \end{equation}
    where $g$ is the coupling constant. We can define the
    't Hooft's coupling constant as $\lambda = g^2 N$,
    where $N$ parameterizes the gauge group U($N$).
    The Hamiltonian reads
    \[ H = \{Q, Q^{\dagger} \} = H_B + H_F,\]
    \begin{eqnarray}
    H_B & = & Tr[\ad a + g(\add a + \ad a^2) + g^2 \add a^2], \\
    H_F & = & Tr[\fd f + g(\fd f (\ad + a) + \fd(\ad +a)f) \\
        & + & g^2(\fd a f \ad + \fd a \ad f + \fd f \ad a + \fd \ad
        f a)]. \nonumber
    \end{eqnarray}
    It conserves the fermionic occupation number
    \mbox{$F = Tr[\fd f]$}, so we can analyze our model separately for
    each fixed $F$. The cases $F=0$ and $F=1$ were described in
    Ref\cite{vw1}, whereas the sectors $F=2$ and $F=3$ in
    Ref\cite{vw2}. We will concentrate here exclusively on
    these higher-fermion-number sectors.

    Following the rules of planar calculus, described in detail
    in Ref\cite{vw1}, one can calculate the matrix elements of
    the Hamiltonian in the sectors with two and three fermions.
    We just recall here the explicit results \cite{vw2}.\\

    \textbf{Two fermion sector}\\
    We use the notation for the matrix element:
    \[ H_{n_1,n_2; m_1,m_2} \equiv \ <2,n_1,n_2| H | 2, m_1, m_2>. \]
    Then
    \begin{gather}
    H_{n_1,n_2;n_1,n_2} = \\
    (n_1+n_2+2)(1+\lambda) - \lambda(2 - \delta_{n_1,0} +
    2 \delta_{n_2,n_1+1}) \nonumber
    \end{gather}
    \begin{gather}
    H_{n_1+1,n_2;n_1,n_2} = H_{n_1,n_2;n_1+1,n_2} =
    \sqrt{\lambda}(n_1+2) \\
    H_{n_1,n_2+1;n_1,n_2} = H_{n_1,n_2;n_1,n_2+1} =
    \sqrt{\lambda}(n_2+2) \nonumber
    \end{gather}
    \begin{gather}
    H_{n_1+1,n_2-1;n_1,n_2} = H_{n_1,n_2;n_1+1,n_2-1} =
    \nonumber \\
    2\lambda(1-\delta_{n_2,n_1+1})
    \end{gather}

%    \pagebreak

    \textbf{Three fermion sector}\\
    Similarly, we denote the matrix element by
    \[ H_{n_1,n_2,n_3; m_1,m_2,m_3} \equiv \ <3,n_1,n_2,n_3| H | 3, m_1, m_2,m_3>. \]
    We have
    \begin{gather}
    H_{n_1,n_2,n_3;n_1,n_2,n_3} = \\
    (n_1+n_2+n_3)(1+\lambda) - \lambda(3 - \delta_{n_1,0} -
    \delta_{n_2,0} - \delta_{n_3,0}), \nonumber
    \end{gather}
    \begin{gather}
    H_{n_1+1,n_2,n_3;n_1,n_2,n_3} = H_{n_1,n_2,n_3;n_1+1,n_2,n_3} = \nonumber \\
    \sqrt{\lambda}(n_1+2)\Delta, \nonumber \\
    H_{n_1,n_2+1,n_3;n_1,n_2,n_3} = H_{n_1,n_2,n_3;n_1,n_2+1,n_3} = \nonumber \\
    \sqrt{\lambda}(n_2+2)\Delta,  \\
    H_{n_1,n_2,n_3+1;n_1,n_2,n_3} = H_{n_1,n_2,n_3;n_1,n_2,n_3+1} = \nonumber \\
    \sqrt{\lambda}(n_3+2)\Delta, \nonumber
    \end{gather}
    \begin{gather}
    H_{n_1+1,n_2-1,n_3;n_1,n_2,n_3} = H_{n_1,n_2,n_3;n_1+1,n_2-1,n_3} =
    \lambda \Delta, \nonumber \\
    H_{n_1,n_2+1,n_3-1;n_1,n_2,n_3} = H_{n_1,n_2,n_3;n_1,n_2+1,n_3-1} =
    \lambda \Delta, \nonumber \\
    H_{n_1-1,n_2,n_3+1;n_1,n_2,n_3} = H_{n_1,n_2,n_3;n_1-1,n_2,n_3+1} =
    \lambda \Delta,
    \end{gather}
    where $\Delta$ is defined by
    \begin{equation}
    \Delta = \left\{ \begin{array}{ll}
        \frac{1}{\sqrt{3}} & \textrm{if for the right state
        $n_1=n_2=n_3$},\\
        \sqrt{3} & \textrm{if for the left state $n_1=n_2=n_3$}, \\
        1 & \textrm{otherwise}.
        \end{array} \right. \nonumber
    \end{equation}

    Previous investigations showed the existence of a transition point in the 't
    Hooft's coupling constant $\lambda$ at \mbox{$\lambda_c=1$}. On one hand,
    it appears as a critical slow down of the convergence of eigenenergies as a function
    of $B_{max}$, and on the other, the spectrum becomes continuous,
    whereas it was discrete away from $\lambda_{c}$. It was possible
    to derive the existence of this transition point analytically in the sectors with
    none or one fermion. Numerical results strongly suggest that the
    transition also occurs at $\lambda_c=1$ in the higher-fermion-number
    sectors. The aim of the present paper is to confirm this by new high precision results
    from larger cut-offs calculations.

    \subsection*{High cut-off results \\ and the transition point}

    Our main goal here is to study in detail the location of the
    transition point $\lambda_c$ in the sectors with two or three fermions.
    Since the bases in these sectors are much bigger than the
    ones in lower-fermion-number, one needs another tool for
    more quantitative analysis. We used ARPACK, a Fortran77
    library for spare matrices, to diagonalize our Hamiltonian matrix.
    In this way we were able to reach cut-offs $B_{max} =
    500 \ (110)$, respectively for $F=2 \ (3)$,
    corresponding to the sizes of basis up to $100000$ vectors,
    compared to $B_{max} = 40 \ (30)$ attained in Ref\cite{vw2}.

    \subsubsection*{Two fermion sector}

\begin{figure}[!t]
    \includegraphics[width=0.5\textwidth]{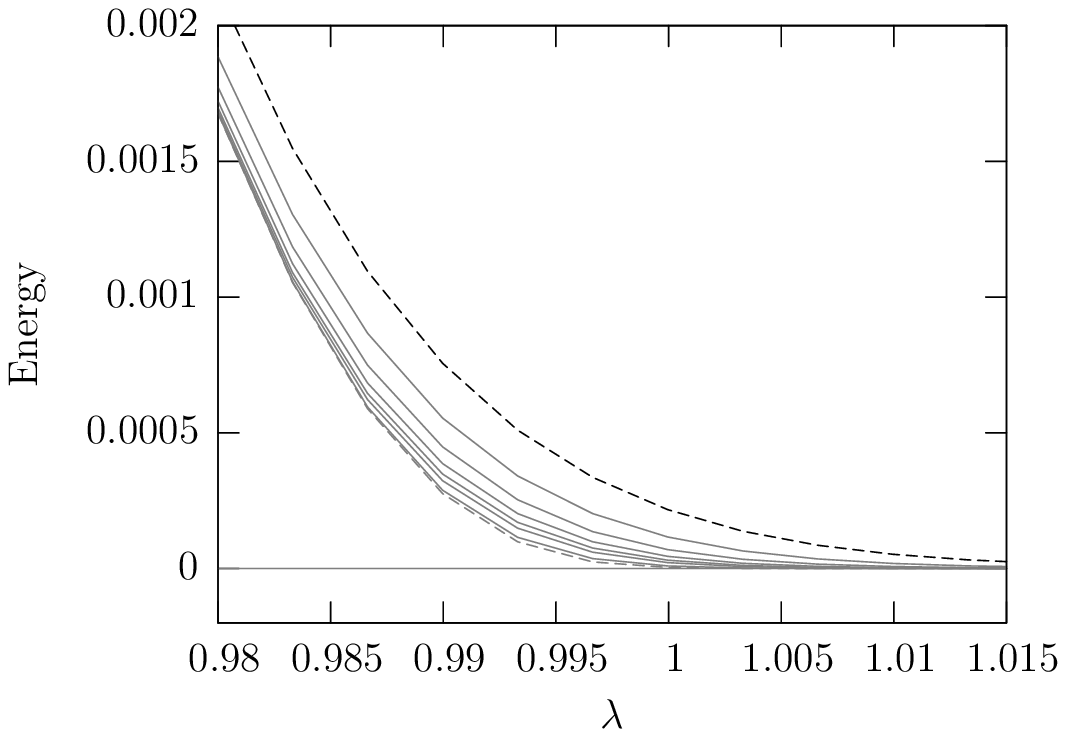}
    \caption{The dependence of the ground state of the sector with
    two fermions
    on the coupling constant $\lambda$ for different cut-offs $B_{max}$. The
    highest dashed curve represents the results for the smallest $B_{max} = 100$,
    whereas the lowest one corresponds to the highest $B_{max} = 400$.
    \label{stan_podf2_a}}
    \end{figure}

\begin{figure}[!t]
    \includegraphics[width=0.5\textwidth]{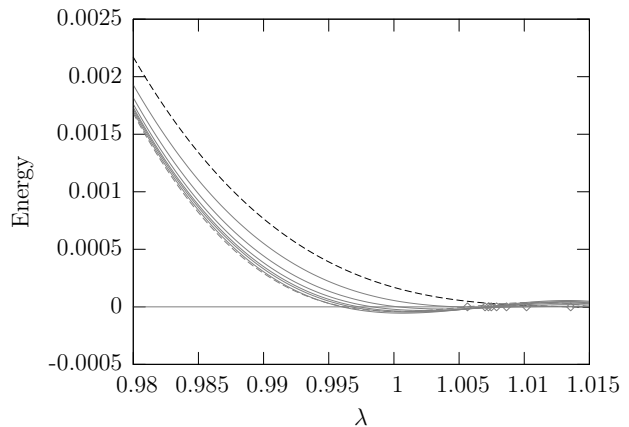}
    \caption{Polynomials fitted to the dependence of the energy of the
    ground state in sector with two fermions on the coupling $\lambda$ for different
    $B_{max}$ as well as their zero points. The
    highest dashed curve represents the results for the smallest $B_{max} = 100$,
    whereas the lowest one corresponds to the highest $B_{max} = 400$.
    \label{stan_podf2_b}}
    \end{figure}

\begin{figure*}[!htb]
    \begin{center}
    \includegraphics[width=0.75\textwidth]{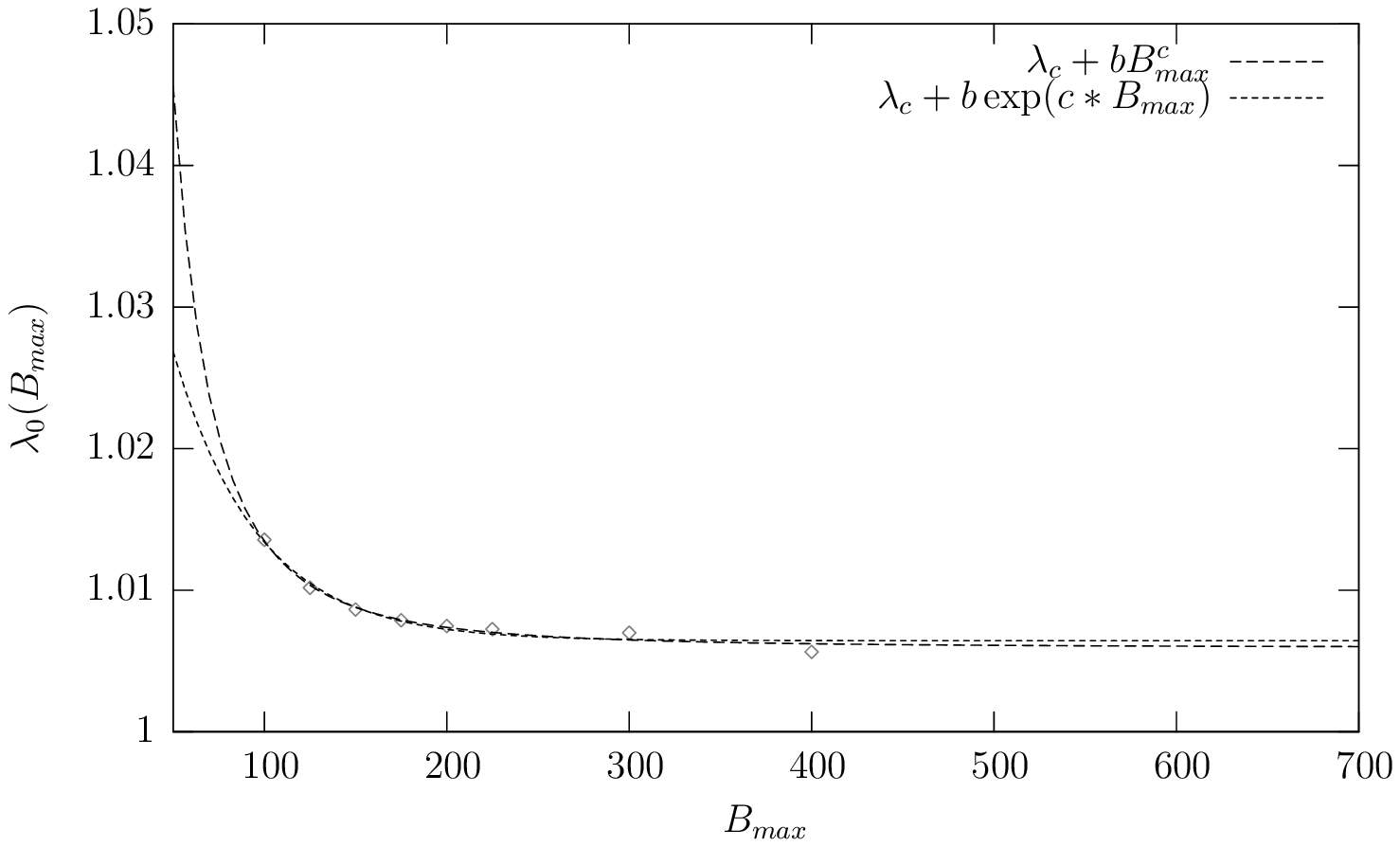}
    \caption{The fitted dependencies of $\lambda_0(B_{max})$ on $B_{max}$. \label{depf2_up}}
    \end{center}
    \end{figure*}

\begin{figure*}[!htb]
    \begin{center}
    \includegraphics[width=0.75\textwidth]{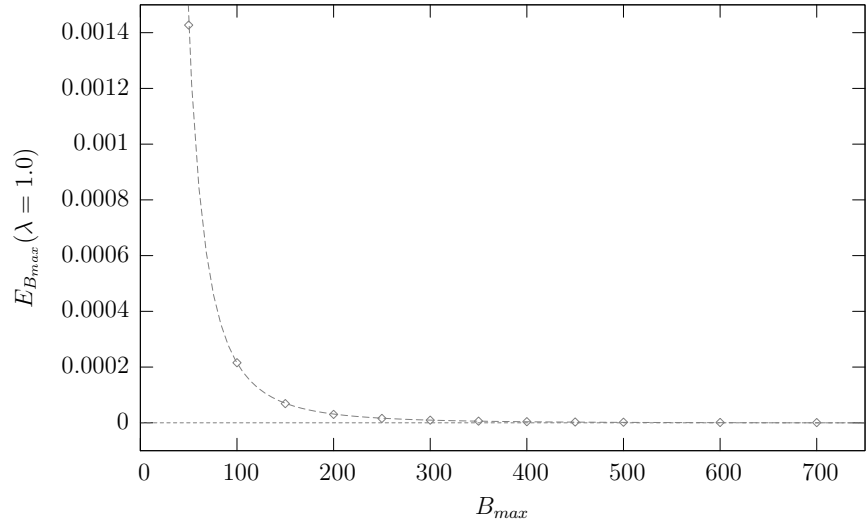}
    \caption{The fitted dependence of $E_{B_{max}}(\lambda=1.0)$ on $B_{max}$. \label{depf2_down}}
    \end{center}
    \end{figure*}

    We will find the transition point by examining the dependence of the energy of the ground state on the coupling constant $\lambda$.
    Figure \ref{stan_podf2_a} shows this energy, which in the following we will call $E_{B_{max}}(\lambda)$, for a given cut-off
    $B_{max}$ and
    in some interval around $\lambda=1$. Suggestions from previous works are
    confirmed. Namely, for $\lambda > \lambda_c$ the ground energy vanishes,
    and thus constitutes one of two SUSY ground states, which are
    present in this sector. For $\lambda <
    \lambda_c$, $E_{B_{max}}(\lambda)$ is non-null and has a nontrivial
    dependence on $\lambda$. The determination of the transition point is carried out by fitting to $E_{B_{max}}(\lambda)$ a
    polynomial in $\lambda$ for several fixed $B_{max}$. This polynomial is chosen to be positive for $\lambda < \lambda_0$
    and equal to zero at $\lambda = \lambda_0$
    \[w(\lambda) = w_1 ( \lambda - \lambda_0) + w_3 ( \lambda -
    \lambda_0)^3 + w_5 ( \lambda - \lambda_0)^5.\]
    The fitted curves, together with the polynomial roots $\lambda_0(B_{max})$, are shown in figure
    \ref{stan_podf2_b}. In order to obtain the value of the physical transition point $\lambda_c$, we extrapolate $\lambda_0(B_{max})$ to the limit $B_{max} \rightarrow
    \infty$. We do this by fitting two types of decreasing functions
    \begin{eqnarray}
    \lambda_0(B_{max}) &=&
    \lambda_c + b (B_{max})^c, \nonumber \\
    \lambda_0(B_{max}) &=& \lambda_c + b
    \exp(c B_{max}). \nonumber
    \end{eqnarray}
    The resulting fits are shown in \mbox{figure \ref{depf2_up}}, whereas \mbox{table \ref{tab fitf2}} contains the values of
    fitted parameters. We can read off the infinite-cut-off limit
    $\lambda_c$ equal to
    \[ \lambda_c = 1.0061 \pm 0.0005, \]
    where the uncertainty is given by the difference between the two values
    of $\lambda_c$ coming from the fits of the two functions.
    \begin{table}[!h]
    \begin{center}
    \begin{tabular}{|c|c|}
    \hline
    Fitted function & Obtained parameters \\
    \hline
    $\lambda_0(B_{max}) = $
    & $\lambda_c = 1.0059$ \\
    $\lambda_c + b (B_{max})^c$ & $b = 471.1$ \\
    & $c = -2.40$\\
    \hline
    $\lambda_0(B_{max}) = $
    & $\lambda_c = 1.0064$ \\
    $\lambda_c + b \exp (c B_{max})$ & $b = 0.060$ \\
    & $c = -0.022$\\
    \hline
    \end{tabular}
    \caption{Numerical values of fitted parameters for the $\lambda_0(B_{max})$ dependence.\label{tab fitf2}}
    \end{center}
    \end{table}
    Finally, we also check the convergence of the ground energy $E_{B_{max}}(\lambda)$
    at the suspected value of transition point \mbox{$\lambda = \lambda_c = 1.0$}.
    To this end, we calculate the extrapolation of the values $E_{B_{max}}(\lambda =
    1.0)$,
    obtained for some specific cut-offs, by fitting a
    function \[E_{B_{max}}(\lambda = 1.0) = E_c + b (B_{max})^c.\]
    The results for the fitted
    parameters are shown in \mbox{table \ref{tab e minf2}}, and the curve
    is plotted in figure \ref{depf2_down}. We can conclude, that
    \[ E_c = -1.08 \ 10^{-6} \pm 4.5 \ 10^{-7}, \] where the error is
    given by the difference between the two results with highest
    cut-off. \\

    To summarize, in this paragraph we have showed, that in the sector with two fermions, the transition point
    occurs for $\lambda_c = 1.0$ and that the
    energy of the ground state at the conjectured transition point \mbox{$\lambda = 1.0$} converges to zero.
    \begin{table}[!h]
    \begin{center}
    \begin{tabular}{|c|c|}
    \hline
    Fitted function & Obtained \\
     & parameters \\
    \hline
    $E_{B_{max}}(\lambda=1.0) = $
    & $E_c = -1.08 * 10^{-6}$ \\
    $E_c + b (B_{max})^c$ & $b = 62$ \\
    & $c = -2.73$ \\
    \hline
    \end{tabular}
    \caption{Numerical values of fitted parameters for the $E_{B_{max}}(\lambda=1.0)$ dependence.\label{tab e minf2}}
    \end{center}
    \end{table}


    \vspace{35cm}

    \subsubsection*{Three fermion sector}

\begin{figure}[!t]
    \includegraphics[width=0.5\textwidth]{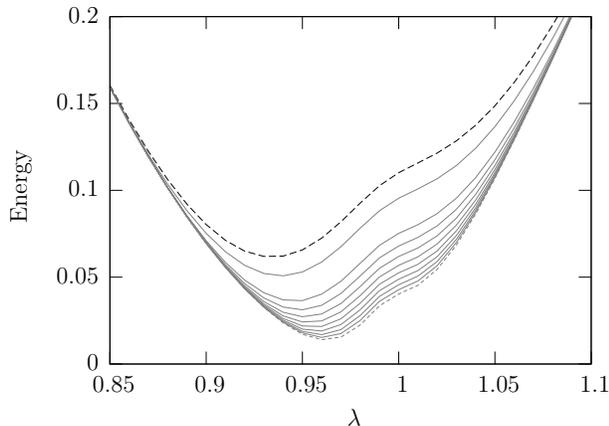}
    \caption{The dependence of the ground state of the sector with
    three fermions
    on the coupling constant $\lambda$ for different cut-offs $B_{max}$. The
    highest dashed curve represents the results for the smallest $B_{max} = 45$,
    whereas the lowest one corresponds to the highest $B_{max} =
    90$. \label{stan_podf3_a}}
    \end{figure}

    The analysis of the transition
    point in this sector follows the lines of the preceding paragraph.
    Similarly, we will examine the dependence of the energy of the
    ground state, called $E_{B_{max}}(\lambda)$, on the coupling $\lambda$. Figure \ref{stan_podf3_a}
    demonstrates the numerical curves for different cut-offs.
    The highest, dashed, curve represents calculations for $B_{max}=
    45$, and the lowest one for $B_{max}=90$. We see that the
    convergence is very good for $\lambda$ away from $\lambda_c$
    i.e. $\lambda < 0.90$ and $\lambda > 1.05$. The transition takes
    place for $\lambda$ between these values, and can be seen on
    this plot as a slow down of the numerical method. Let's denote,
    for each $B_{max}$, the minimal energy of
    the ground state by $E_{min}(B_{max})$
    and its position by $\lambda_{min}(B_{max})$.
    The physical results, i.e. cut-off independent, are thus the limiting
    quantities $E_c$ and $\lambda_c$ such that \mbox{$E_{min}(B_{max}) \rightarrow
    E_c$}
    and \mbox{$\lambda_{min}(B_{max}) \rightarrow \lambda_c$}
    as $B_{max} \rightarrow \infty$. The results from sectors
    $F=0,1$ suggest that
    the energy of all states collapses to zero and we get a
    continuous spectrum at the speculated transition point $\lambda_c =
    1$.

    \begin{figure}[!t]
    \includegraphics[width=0.5\textwidth]{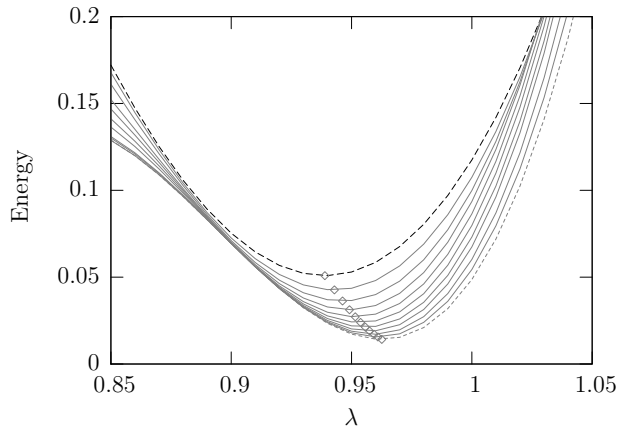}
    \caption{Polynomials fitted to the dependence of the energy of the
    ground state in sector with three fermions on the coupling $\lambda$ for different
    $B_{max}$. The highest dashed curve corresponds to $B_{max}=45$
    and the lowest one to $B_{max}=90$. The calculated minima are
    shown as well. \label{stan_podf3_b}}
    \end{figure}

\begin{figure*}[!htb]
    \begin{center}
    \includegraphics[width=0.75\textwidth]{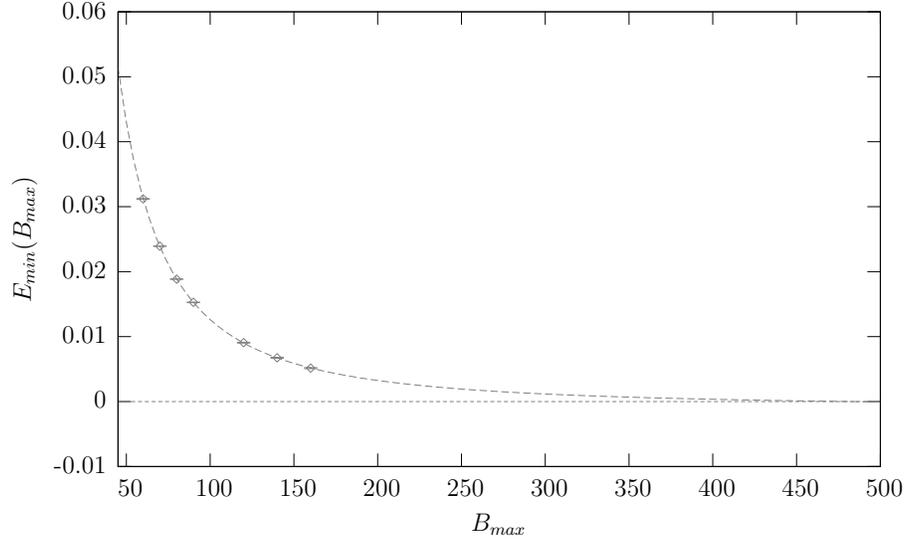}
    \caption{The fitted dependence of $E_{min}(B_{max})$ on $B_{max}$. \label{dep_up}}
    \end{center}
    \end{figure*}

\begin{figure*}[!htb]
    \begin{center}
    \includegraphics[width=0.75\textwidth]{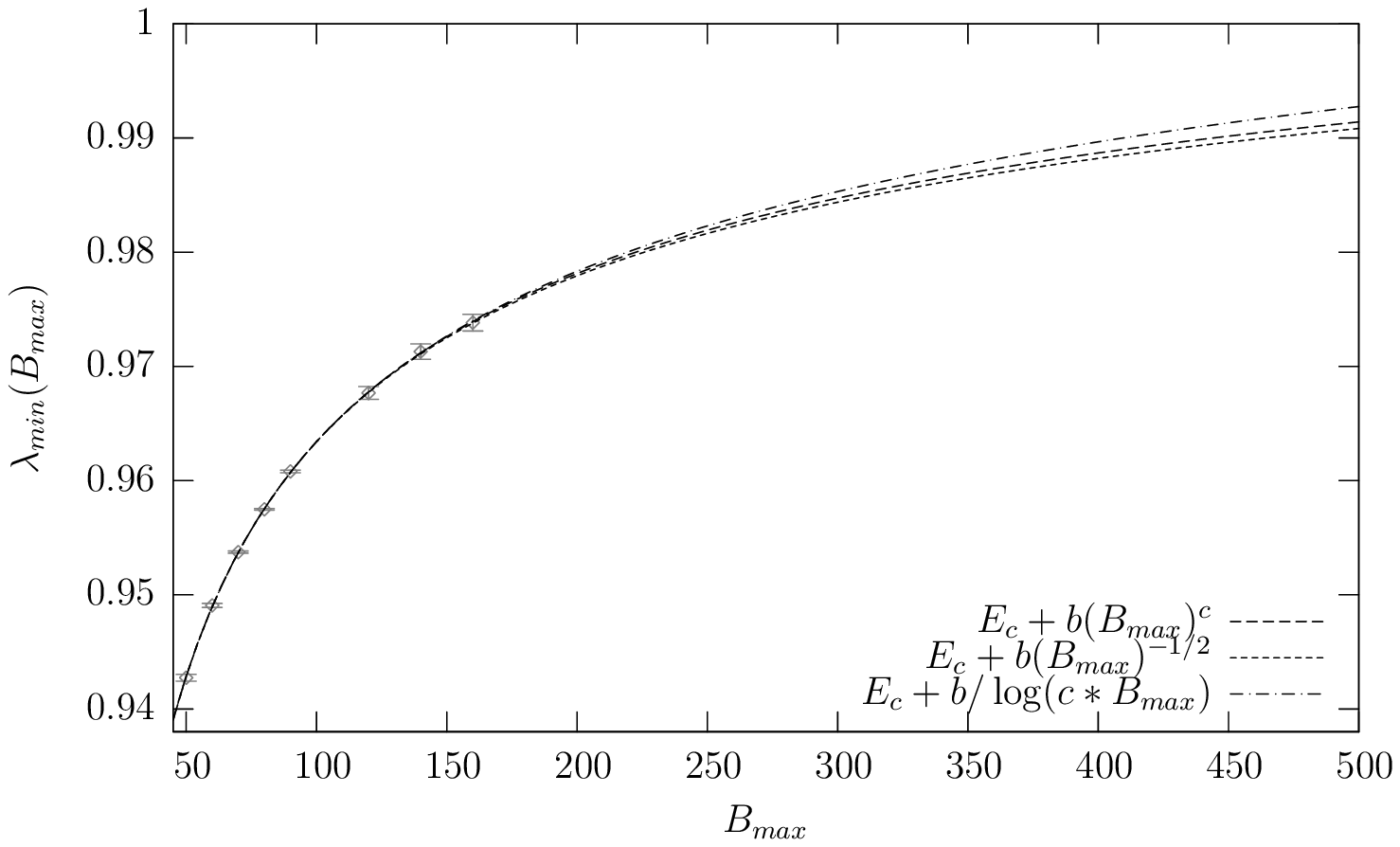}
    \caption{The fitted dependencies of $\lambda_{min}(B_{max})$ on $B_{max}$. \label{dep_down}}
    \end{center}
    \end{figure*}

    We determine $\lambda_{min}(B_{max})$ and $E_{min}(B_{max})$
    by two methods. First of them consists in fitting a fourth
    order polynomial,
    \[ w(\lambda) = w_0 + w_1 \lambda + w_2 \lambda^2 + w_4 \lambda^4, \]
    to $E_{B_{max}}(\lambda)$ for each $B_{max}$. Then, $\lambda_{min}(B_{max})$ and
    $E_{min}(B_{max})$ are calculated analytically given the
    fitted parameters. The fitted curves are shown in \mbox{figure
    \ref{stan_podf3_b}}, together with the calculated minima. The second
    method uses the cubic spline to transform $E_{B_{max}}(\lambda)$
    into a continuous curve. The approximated values of the minima are then found
    numerically by bracketing. The differences between the results
    coming from these two methods will be later used as an estimate of the
    uncertainty of the calculated quantities. To get $E_c$ we
    extrapolate $E_{min}(B_{max})$ to \mbox{$B_{max} \rightarrow
    \infty$}, and to this end, we fit a polynomial function
    \[
    E_{min}(B_{max}) = E_c + b (B_{max})^c.
    \]
    The obtained fit is shown in figure \ref{dep_up}, whereas the values of
    parameters are presented in table \ref{tab e min}. We thus have
    \[ E_c = -0.00094 \pm 0.00021.\]
    \begin{table}[!h]
    \begin{center}
    \begin{tabular}{|c|c|}
    \hline
    Fitted function & Obtained \\
     & parameters \\
    \hline
    $E_{min}(B_{max}) = $
    & $E_c = -0.00094 \pm 0.00021$ \\
    $E_c + b (B_{max})^c$ & $b = 33.2 \pm 2.4$ \\
    & $c = -1.694 \pm 0.019$ \\
    \hline
    \end{tabular}
    \caption{Numerical values of fitted parameters for the $E_{min}(B_{max})$ dependence.\label{tab e min}}
    \end{center}
    \end{table}
    In order to extrapolate $\lambda_{min}(B_{max})$ we fit three
    slowly growing functions: \begin{itemize}
    \item \mbox{$\lambda_{min}(B_{max})
    \sim (B_{max})^c$},
    \item \mbox{$\lambda_{min}(B_{max}) \sim (B_{max})^{-1/2}$}
    \item \mbox{$\lambda_{min}(B_{max})
    \sim \ln(B_{max})^{-1}$}.
    \end{itemize}
    Table \ref{tab lambda min} contains the
    obtained values of the fitted parameters, and the fitted curves
    are shown in figure \ref{dep_down}.
    \begin{table}[!t]
    \begin{center}
    \begin{tabular}{|c|c|}
    \hline
    Fitted functions & Obtained \\%& Red. \\
     & parameters \\%& $\chi^2$ \\
    \hline
    $\lambda_{min}(B_{max}) = $
    & $\lambda_c = 1.0155 \pm 0.0053$ \\%& $0.40$ \\
    $\lambda_c + b (B_{max})^c$ & $b = -0.474 \pm 0.045$ \\%& \\
    & $c = -0.480 \pm 0.043$ \\%& \\
    \hline
    $\lambda_{min}(B_{max}) = $ & $\lambda_c = 1.01304 \pm 0.00038$ \\%& $0.35$ \\
    $\lambda_c + b (B_{max})^{-1/2}$ & $b = -0.4964 \pm 0.0033$ \\%& \\
    \hline
    $\lambda_{min}(B_{max}) = $ & $\lambda_c = 1.073 \pm 0.010$ \\%& $0.38$ \\
    $\lambda_c + b \ln(c*B_{max})^{-1}$ & $b = -0.478 \pm 0.084$ \\%& \\
    & $c = 0.80 \pm 0.28$ \\%& \\
    \hline
    \end{tabular}
    \caption{Numerical values of fitted parameters for the $\lambda_{min}(B_{max})$ dependencies.\label{tab lambda min}}
    \end{center}
    \end{table}
    Eventually, we can assume that the value of the constant coefficient $\lambda_c$ is
    equal to the mean of the values obtained from the three fits,
    and its error is the standard deviation. Therefore
    \[
    \lambda_c = 1.034 \pm 0.016.
    \]
    One also notes, that the general fit of a power function gave
    approximately the same results as the fit of the inverse of the
    square root.\\

    As a conclusion of this section we recapitulate our results for the sector with three fermions.
    Namely, we showed that the transition point occurs at $\lambda_c
    = 1.0$ and that at this value of coupling constant the ground
    energy converges to zero.

    \subsection*{Discussion and conclusions}

    In this paper we used high precision numerical results in order
    to check the transition point in the 't Hooft's coupling
    constant $\lambda$ in the Veneziano-Wosiek model. We investigated the
    sectors with two and three fermions. By fitting some specific
    functions we extrapolated from the numerical data the physical,
    i.e. cut-off independent, values of the transition point and
    ground energy at $\lambda = 1.0$. We confirmed that in both sectors this
    transition point occurs nearly at $\lambda = \lambda_c = 1.0$, and that
    the ground energy at this value of coupling constant converges almost to
    zero. The uncertainty given with these results is not a true statistic
    error since it was not calculated from any statistical ensemble.
    It should be only interpreted as an indication of the real
    uncertainty.

    \subsubsection*{Acknowledgments}

    I would like to thank prof. J. Wosiek for suggestions and help with
    this paper. This work is partially supported by the grant No. P03B
    024 27 (2004-2007) of the Polish Ministry of Education and Science.

\end{document}